# VARIANCE IN SYSTEM DYNAMICS AND AGENT BASED MODELLING USING THE SIR MODEL OF INFECTIOUS DISEASE

Aslam Ahmed, Julie Greensmith, Uwe Aickelin
Intelligent Modelling and Analysis Research Group
University of Nottingham
Nottingham, NG8 1BB, United Kingdom
E-mail:{awa, jqg, uxa} @cs.nott.ac.uk

**ABSTRACT**

Classical deterministic simulations of epidemiological processes, such as those based on System Dynamics, produce a single result based on a fixed set of input parameters with no variance between simulations. Input parameters are subsequently modified on these simulations, using Monte-Carlo methods, to understand how changes in the input parameters affect the spread of results for the simulation. Agent Based simulations are able to produce different output results on each run based on knowledge of the local interactions of the underlying agents and without making any changes to the input parameters. In this paper we compare the influence and effect of variation within these two distinct simulation paradigms and show that the Agent Based simulation of the epidemiological SIR (Susceptible, Infectious, and Recovered) model is more effective at capturing the natural variation within SIR compared to an equivalent model using System Dynamics with Monte-Carlo simulation. To demonstrate this effect, the SIR model is implemented using both System Dynamics (with Monte-Carlo simulation) and Agent Based Modelling based on previously published empirical data.

**INTRODUCTION**

Models of infectious diseases can be useful for understanding the spread of infection of the diseases within a population over time. However, within a given population, diseases can spread at different rates over time due to the natural random nature of contact between individuals in the population. If a simulation can incorporate this kind variation, the extra information can be used to determine the spread of uptake of infection in worst case and best case scenarios for a given population.

Currently, for classical System Dynamics (SD) models (Forrester, 1961) based on ordinary differential equations, the random contact between individuals is aggregated to fixed rates of contact and the output has no variation. Assuming the same parameter values are supplied to the System Dynamics simulation, on each run, the same results are produced. Subsequently, in order to understand the spread of output values, the simulations are repeated with different input parameters by applying Monte Carlo simulation (Stan, 1987). In this approach, multiple experiments are performed and the parameter values taken from a probability density function representing the input parameter range. In Agent Based models (ABM), uncertainty or variance can be inherent within the model so that the simulations from the models produce non-deterministic results directly without input parameter variation.

In this paper, the two approaches are examined by generating an SD model with Monte-Carlo and an ABM and comparing the spread of output values against published data for a defined population. Simulations from modelling paradigms such as Agent Based Modelling, which can include variance, help to bridge the gap between raw data and simulation data and also help answer the issue of validation in simulation - assessing the degree to which a model is an accurate representation of the real world (Oberkampf et al., 2002). Both System Dynamics and Agent Based models are able to capture overall variance but unlike simulations from SD models, a single simulation run from an Agent Based Model is able to capture the 'typical' outcome from a single simulation experiment.

Unlike System Dynamics which uses a top-down approach to model the system as a whole, in Agent Based simulations, the system is 'brought about' by carrying out the lower level interactions between the agents. For this reason, ABM is beginning to be used in a range of fields including biological simulations and social sciences representing people as interacting agents in environments (Zellner, 2008)(Siebers et al., 2010).

**VARIANCE IN EPIDEMIOLOGICAL SYSTEMS**

Early Mathematical models for epidemiology such as those by Bernoulli in 1766 (Dietz and Heesterbeek, 2002) were useful deterministic models that could be used to determine 'what-if' scenarios such as the change in life expectancy following the introduction of inoculation against smallpox. Further models followed including those for SIR proposed by Kermack and McKendrick which were stochastic (McKendrick, 1926) and deterministic (Kermack and McKendrick, 1927).

Early opinions for mathematical modelling of epidemic models were that deterministic models gave an average outcome of a corresponding stochastic model and that for large populations, it was the average that mattered. A more recent understanding is that both deterministic and stochastic models have their strengths and weaknesses.

Traditional deterministic models of epidemiology assume heterogeneity of mixing. It is assumed that individuals have the same rate of contact with others and recovery from infection takes the same time. In reality, contact rate is affected by transport networks and individual lifestyle and recovery from infection can depend on age and other factors so these are not take into account. Sometimes this data is difficult to ascertain but in smaller population sizes it may be possible to obtain this information and build a model that is a closer representation to reality.

One of the underlying reasons why epidemiological systems exhibit variation is due to the complex way that the individuals in a population have contact with each other. Infection levels can coincide with transport networks such as road and rail so individuals in areas with high levels of such transport links are more susceptible to catching infection. At a much lower level, random variation exists due to Brownian motion of the interaction between molecules (Gaspard, 2005).

## AGENT BASED MODELLING AND SYSTEM DYNAMICS

Agent Based Modelling (Macal and North, 2008) is a more recent addition to the set of tools for simulations compared to classical mathematical models. ABM uses agents, which are discrete autonomous entities containing characteristics and rules which govern their behaviour and interaction with other agents. Agents can be programmed to adapt and learn from previous interactions. An ABM can have closer affinity with the system being modelled as the notable entities and their significant properties can be captured making the simulation more intuitive and closely resembling the real system. In System Dynamics (Sterman, 2004), complex non-linear systems are represented using feedback loops and delays by creating stocks which represent quantities over time, flows which measure the transition from stock to stock and factors which influence the values of the flows.

## VARIANCE EXPERIMENT FOR ABM AND SD

In order to show how the variance differs in ABM and SD two models are built. One using SD with Monte Carlo simulation (to drive variation) and one using ABM which has variance built into the design. A basic Susceptible-Infected-Recovered (SIR) model, originally proposed by Kermack and McKendrick is used for both types of modelling paradigms. The SIR model is a simple but effective model of infection that has been used to represent a wide range of epidemics including influenza, tuberculosis, chicken pox, rubella and measles (Enns, 2011).

In the basic SIR model, each person is in a state of:
- Susceptible
- Infected
- Recovered

A person who is susceptible has never been infected. As soon as they are infected by way of contact with an infected individual, they are set to the Infected state. After a period of time, during which the immune system is able to recover from the infection, an individual moves from the Infected to the Recovered state. Once in a Recovered state, the individual is immune to further infections.

### SIR Data

The experimental data used in this paper is obtained from the Russian Influenza epidemic in Sweden between 1889 and 1890 (Skog, 2008). After the outbreak of Russian Flu in Sweden, questionnaires were sent to Swedish physicians to determine ascertain information about the flu in their region. Answers were received by 398 physicians for over 32,600 individuals. The information returned from the postcard included the following for each district:
- Date of first influenza case detected
- Peak of epidemic
- Percentage affected

The information returned from the questionnaire included the following for each household:
- Number of infected persons
- Gender
- Age

Some of the important findings are used as the parameter values. This included the duration of the disease which was found to be between 2.3 and 9.4 days and affected 61% of the population. For the purpose of the experiment, a single area, Österlövsta, was chosen for analysis. This area has a profile which shows a typical raise and decline of infection population counts over a period of 15 weeks. The complete data for Sweden can also be used but this will extend the ABM simulation time. The data obtained is shown in Figure 1.

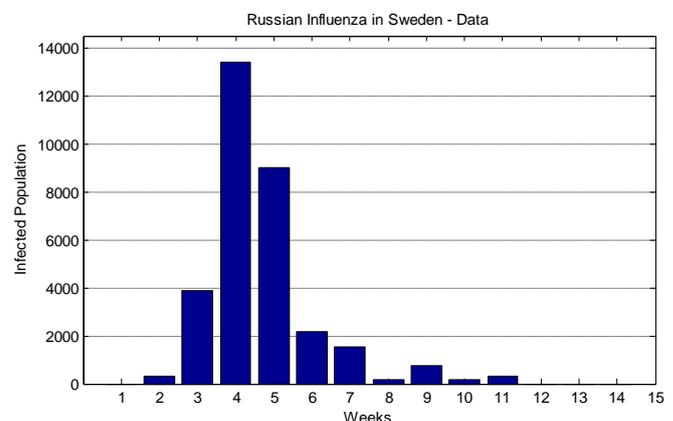

Figure 1: Russian Influenza in Sweden – Data

Data was taken from parishes surrounding one of the regions, Österlövsta, based on original work by Linroth over a period of 15 weeks. The total population size is 52910 and is used in the experiments. The other parameters for the models are based on the main findings of the study with the illness duration set to 4.2 and the probability of infection set to 0.065. This produced a best fit for the selected region of Österlövsta.

**System Dynamics Model**

The SD model is based on the original SIR model proposed by Kermack and Kendrick (Kermack and McKendrick, 1927). The model captures the spread of a contagious disease in a closed population over time. Three coupled, ordinary differential equations are used to represent the rate of change of the three different states of the people in a given population. The equations in the model are shown which the rate of change for each of the components of SIR.

$$\frac{dS}{dt} = -aSI$$
$$\frac{dI}{dt} = aSI - bI$$
$$\frac{dR}{dt} = bI \qquad (1)$$

The meaning of the parameters is shown in Table 1.

Table 1: Parameter Description for SIR Equations

| Parameter | Description |
| --- | --- |
| a | Infection rate |
| b | Recovery rate |
| S | Susceptible population |
| I | Infected population |
| R | Recovered population |

From the equations, a System Dynamics model is built in AnyLogic with stocks labelled as Susceptible, Infectious and Recovered and flows labelled as Infection Rate and Recovery Rate as shown in Figure 2.

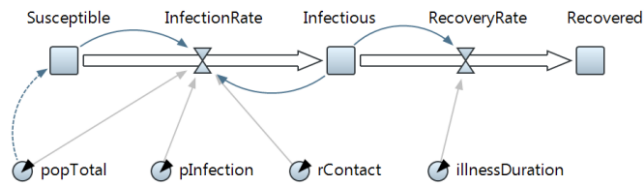

**Figure 2: Stocks and Flows for the System Dynamics Model**

The system dynamics model produces the results shown in Figure 3.

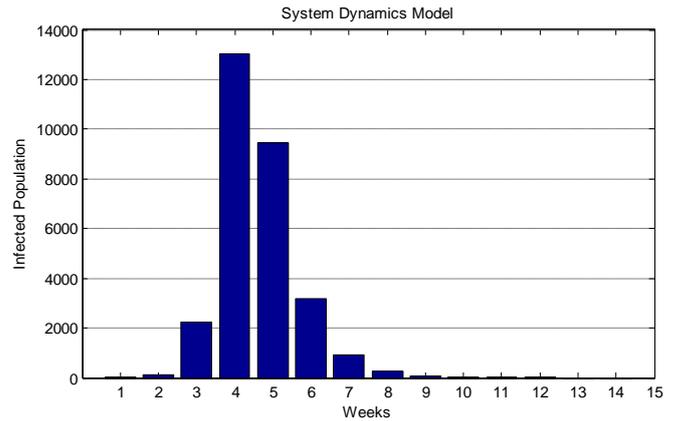

Figure 3: Results for the System Dynamics Model

**Agent Based Model**

In the ABM model, each person is in a state of **Susceptible**, **Infectious** or **Recovered**. A person in the **Susceptible** state moves to the **Infectious** State on receipt of a message representing the transfer of the infection from one person to another.

The infection is passed from one agent to another randomly connected agent in the network to another at a fixed contact rate and an individual recovers from the infection using a recovery rate. A state chart is used to model the state of the agent as shown in Figure 4.

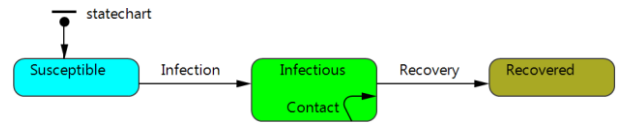

Figure 4: State Chart for the Agent Based Model

As per the System Dynamics model, a population count of 52910 is used in the model based on the Russian Influenza epidemic in Sweden. Individuals (agents) are connected using a small world (Watts and Strogatz, 1998) network topology. This is chosen as it represented a suitable route of transmission of the infection with many close connections in the network coupled with distant connections. The distant connections may be perceived as transportation links such as those by rail or sea. The small world network has been used in epidemiology (da Gama and Nunes 2006) with a number of studies using it as part of the models (Boots and Sasaki, 1999). A single randomly connected agent is chosen to kick-start the spread of infection.

A feasibility study is carried out for the modelling software and AnyLogic by XJ Technologies chosen as a suitable choice for modelling SIR in System Dynamics and Agent Based Modelling. One of the features of AnyLogic is that it has inherent support for combining different modelling paradigms into a single model.

In total, 100 experiments are carried out in AnyLogic. The experiments are carried out on a PC running Windows 7 with 3GB memory and an Intel Core 2 P8700 microprocessor. Output from the experiments is imported into MatLab to generate the box plots. The result for the ABM is shown in Figure 5.

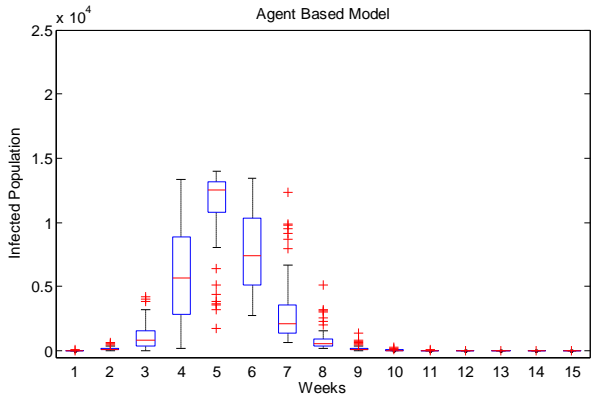

Figure 5: Agent Based Model Results

As per the SD model, the AB model is also validated against the data from the Influenza epidemic. The ABM simulation takes a total of 13 hours to complete.

**SD Monte Carlo Simulation**

The Monte Carlo simulations are used to determine how infected population counts change when the input parameters to the SD model are varied. Monte Carlo simulation uses repetitions of random sampling of the input parameters to determine the result. The randomness is applied 'outside' of the internal workings of the system as it is the parameters to the system being sampled.

One of the limitations of using the Monte Carlo method applied to simulations is the time taken to perform the simulation over a very large number of iterations. Therefore in areas such as Probability Sensitivity Analysis, the Monte Carlo solution is not always a viable method for complex models such as those for healthcare, involving thousands of patients (O'Hagan et al., 2007).

Monte Carlo simulations are carried out using the SD model to see the effect of varying each parameter and the effect of varying all parameters. In total, 100 simulations are carried out for each experiment to match the ABM. Parameter variation is carried out by randomly selecting values for each of the parameters taken from a standard normal distribution based on the mean value.

The following Monte Carlo simulations are carried out:

- Illness Duration variation
- Contact rate variation
- Infection rate variation
- Illness duration, contact and infection rate variation

Each SD experiment, comprising 100 simulations, takes a total of 9 seconds. The box plot for the SD Monte Carlo model with illness duration variation is shown in Figure 6. The infected population peaks at 21,442.

In the case where the contact rate is varied, the result is shown in Figure 7. In this case, the inter-quartile range is larger than the simulation where the illness rate is varied and clearly visible in weeks 3 to 7 inclusive.

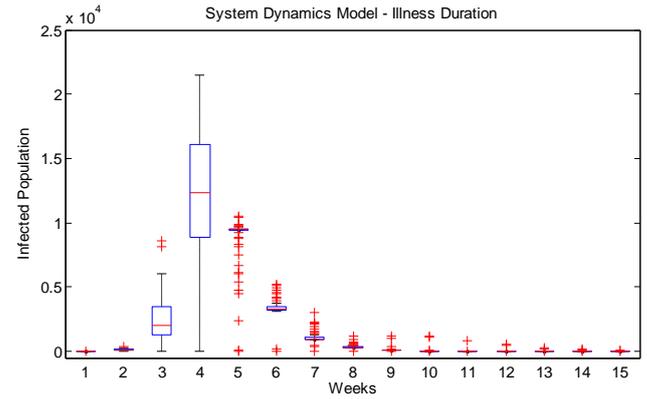

Figure 6: System Dynamics Model - Illness Duration Variation

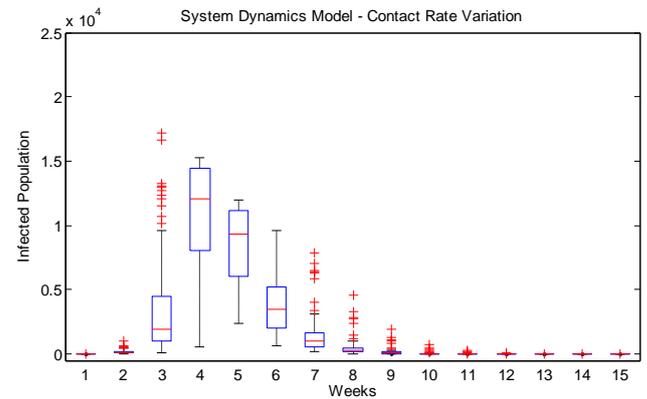

Figure 7: System Dynamics Model - Contact Rate Variation

The SD Monte Carlo model where the infection rate is varied is shown in the box plot in Figure 8.

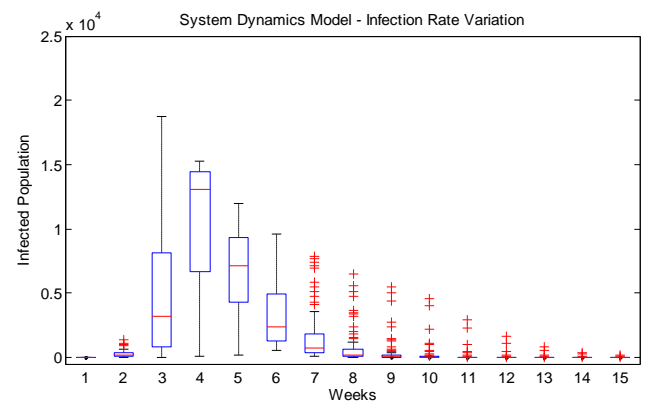

Figure 8: System Dynamics Model - Infection Rate Variation

The box plot for the experiment where multiple parameters are varied is shown in Figure 9. The results show that with multiple parameters being varied the infected population peaks at 24,725 which is a substantial increase compared with the SD version without Monte Carlo simulation which peaks at 13,025. Therefore compared with experiments where variations of contact rate and infection rate are altered to introduce randomness, the variation of multiple parameters has the undesired effect of scaling up the infected population counts.

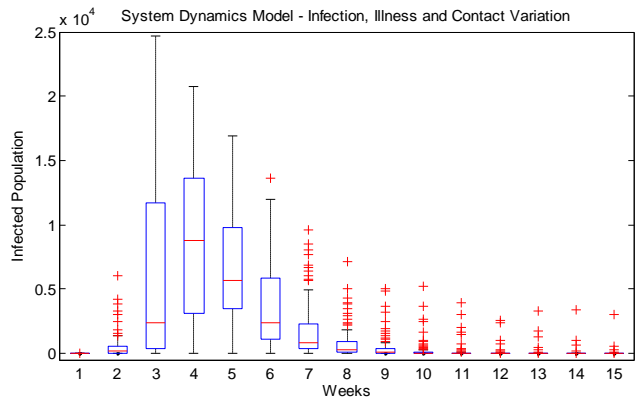

Figure 9: System Dynamics Model - Infection, Illness and Contact Rate Variation

**Validation against Influenza data**

The Wilcoxon signed rank test (Wilcoxon, 1945) is used to compare the simulation results against the Influenza data. This is a non-parametric paired test that tests the null hypothesis that the means for the two data sets are the same versus the means from the two data sets differ.

The SD result without any Monte Carlo simulation is compared directly against the Influenza data. For ABM and SD with Monte Carlo, the median values for each experiment are obtained for each week.

The Wilcoxon rank sum test for the experiment is calculated using MatLab version R2010b. The results are summarized in Table 2. A 5% significance level is used.

Table 2: Wilcoxon Signed Rank Test for experiments

| Simulation | p Value |
|---|---|
| SD | 0.3013 |
| ABM | 0.4648 |
| SD – Vary illness duration | 0.2661 |
| SD – Vary contact rate | 0.2036 |
| SD – Vary infection rate | 0.0244 |
| SD – Vary illness, contact, infection rate | 0.0269 |

The h value for the tests is 0 and the p values of 0.0244 and 0.0269 indicate that the null hypothesis can be rejected for the experiment where infection rate is varied and for the version in which combined parameters are varied.

The Wilcoxon rank sum tests show that the SD without Monte Carlo and the ABM has equivalent overall fits with the experimental data. The ABM experiment, with natural variation between different simulations, due to the contacts between the agents, is in agreement with the Influenza data.

When the Monte Carlo simulation is applied to the SD model, the overall results of the simulation are in agreement with the Influenza data for illness duration variation and contact variation but for variations of infection rate and the combined variation the results are no longer in agreement. The last, combined Monte Carlo simulation, has the overall effect of scaling up the median values overall.

**Variance in ABM and SD Experiments**

Variance for each of the Monte Carlo experiments are taken from the box plots and compared against the variance of the ABM experiment. In order to compare the variances, the inter-quartile range (IQR) is calculated using the MatLab for the ABM experiment and SD Monte Carlo experiments. The IQR for the ABM is shown in Figure 10. The ABM experiment produces a broadly symmetric result reflecting variations of the uptake of the infection which occurs at different times in the simulations but producing the same shape of the infection curve.

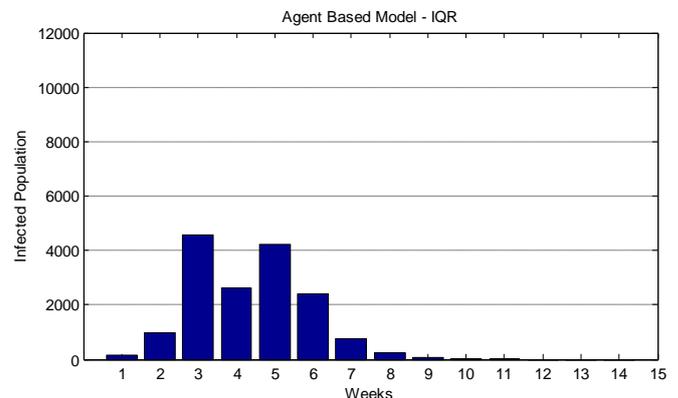

Figure 10: Agent Based Model - IQR

The IQR for the SD Monte Carlo simulation with variations in the illness duration is shown in Figure 11. The chart shows that there is less variation at the height or peak of infections. This is because the variation is created by the random connections that the individuals have in the simulations and critical parameters such as infection rate and illness are kept constant.

The results show that ABM is able to maintain stable peak infection values whilst at the same time exhibiting the type of randomness one may expect between different populations.

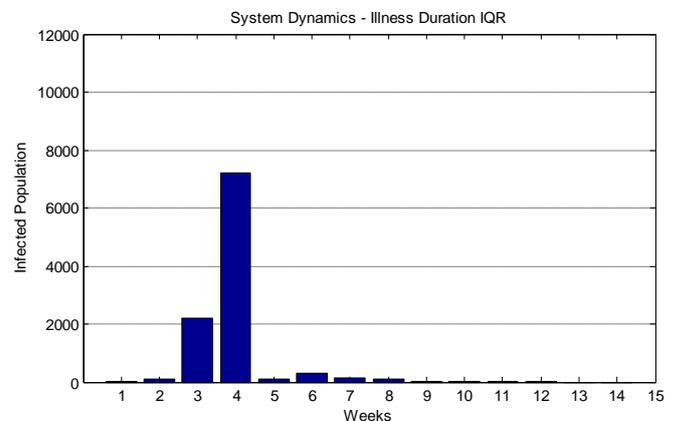

Figure 11: System Dynamics Model – Illness Duration IQR

The chart in Figure 12 shows the IQR where the Contact

Rate is varied in the SD Monte Carlo simulation.

The chart shows that the counts at the height of infection vary significantly between simulations compared to the ABM. In the ABM experiment, the contact rate is constant among the simulations and therefore in those simulations there is less difference of the counts at the height of infection.

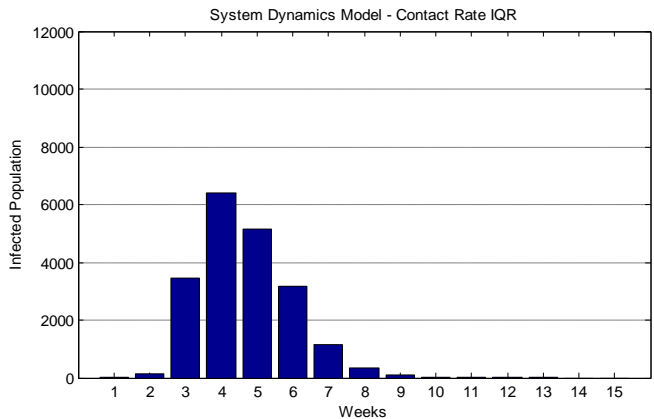

Figure 12: System Dynamics Model – Contact Rate IQR

The chart for the IQR for the infection rate variation for the SD Monte Carlo simulation is shown in Figure 13.

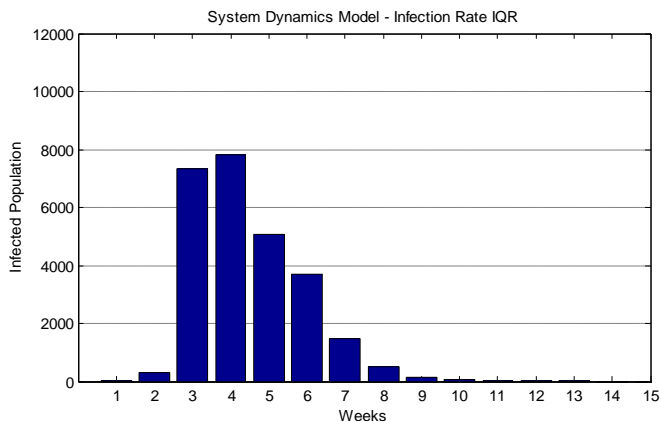

Figure 13: System Dynamics Model - Infection Rate IQR

The chart for the IQR in the case where multiple parameters are varied in the SD Monte Carlo simulation is shown in Figure 14.

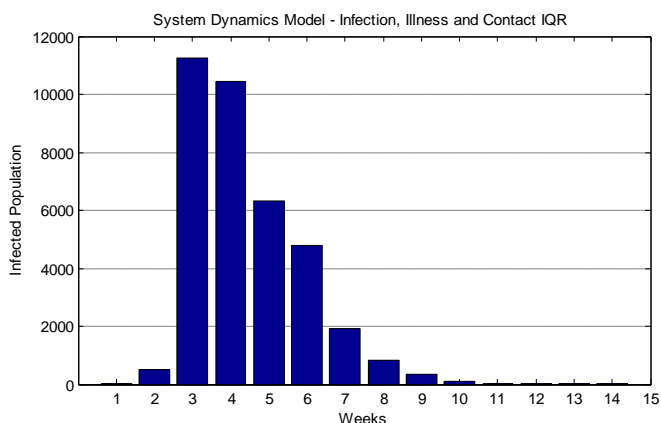

Figure 14: System Dynamics Model - Infection, Illness and Contact IQR

Unlike the Monte Carlo simulations where a single parameter is varied, in this case, with multiple parameter variations, there is an overall significant increase in the variation. Table 3 shows the total variation (the sum of IQR values) for the different experiments.

Table 3: Total variation for the SIR experiments

| Simulation | Total Variation |
|---|---|
| ABM | 16050 |
| SD – Vary illness duration | 10223 |
| SD – Vary contact rate | 19956 |
| SD – Vary infection rate | 26452 |
| SD – Vary illness, contact, infection rate | 36569 |

The least total variation for the simulation is obtained for the Monte Carlo experiment where the illness duration is varied. The Monte Carlo experiment where combined parameters are varied has more than twice the total variation of the ABM experiment.

Results from the ABM simulation showed that the overall peak total infection remained stable between simulations. The shape of the output curve for each simulation has a closer fit with the curve for the empirical data. The ABM simulations only differed with the initial delay before the uptake of infection which may also arise due to natural variation of the contact rate between individuals and their transport networks.

In contrast, for the SD model, the effect on the variation of the parameters has the effect of altering the rate at which the infection spread within the population.

**DISCUSSION AND CONCLUSION**

Although variations of SD models exist which are able to integrate random elements (Tuckwell and Williams, 2007)( Volz and Meyers, 2007) they produce a different kind of variation compared to ABM. Whereas in stochastic models there is a random element applied to the equations, in ABM the randomness is inherent and more natural, following the rules of the underlying system being modelled.

The ABM and SD experiments for the SIR data show that ABM is able to capture natural variation without recourse to modification of any parameters for a simulation. The classic SD model has no variation. The SD with Monte Carlo simulation has variation but it is very sensitive to parameter changes and in the case where multiple parameters are varied, it produces variation and infected population counts which no longer match up against the experimental data. Therefore an ABM of SIR with built-in randomness is able to capture the natural variation in SIR better than a classic SD model with Monte Carlo simulation. The source of variation for the ABM is the contact between the agents between the different experiments.

Several comparative studies between ABM and SD have been undertaken (Jaffry and Treur 2008). Some notable discussions in these studies include the issue of computing power and control. As concluded in this study, the ABM is computationally expensive compared to classical mathematical models although this may be overcome in future by highly parallel computing architectures (Tang et al., 2008). The ABM does however provide more control over how individual agents interact and could be viewed as a more 'faithful' interpretation of the processes being modelled.

As the ABM is built using autonomous individuals, it could be extended to include connections between individuals across different regions to understand the effect of disrupting the spread of the epidemic by shutting down major transport links for example. Further work could include the effect of the use of different network topologies.

Stochastic computer simulation is being used for biochemical network dynamics (Wilkinson, 2009).

The use of ABM with its inherent and intuitive representation of natural variation and interaction among components can help to bridge the gap between computer simulation and biological systems and provide insight of how local level interactions bring about global system outcomes.

## AUTHOR BIOGRAPHIES

**Aslam Ahmed** is a PhD student in the Intelligent Modelling and Analysis Group at the University of Nottingham. His interests include the role of System Dynamics and Agent Based Modelling in specific areas of the immune system.

**Julie Greensmith** is a lecturer in the School of Computer Science at the University of Nottingham. Her interests include artificial immune systems and a wide range of application areas including computer security, affective computing and wearable biosensing.

**Uwe Aickelin** is an EPSRC Advanced Research Fellow and Professor of Computer Science at The University of Nottingham, where he is also the Director of Research in the School of Computer Science and leads one of its four research groups: Intelligent Modelling & Analysis (IMA).